\documentclass[nologo,url,11pt,a4paper]{ETHpaper}
\usepackage{tabularx}                  
\usepackage{MnSymbol}
\usepackage{booktabs}
\usepackage[bf,format=plain]{caption} 
\usepackage{hhline}
\usepackage[table,dvipsnames]{xcolor}
\usepackage{graphicx}                  
\usepackage[english]{babel}            
\usepackage[square,numbers,compress]{natbib}
\usepackage{setspace}
\usepackage{endnotes}

\newcommand{\mean}[1]{\left\langle #1 \right\rangle}

\begin{document}

\title{The network structure of city-firm relations}
\titlealternative{The network structure of city-firm relations}
\author{Antonios Garas$^{1}$, C\'eline Rozenblat$^{2}$ and Frank Schweitzer$^{1}$}
\authoralternative{A. Garas, C. Rozenblat \& F. Schweitzer}
\address{$^{1}$Chair of Systems Design, ETH Zurich, Weinbergstrasse 58, 8092 Zurich, Switzerland\\
$^{2}$Geography Institute, Faculty of Geosciences, University of Lausanne,\\ 1015 Lausanne, Switzerland}


\www{\url{http://www.sg.ethz.ch}}

\makeframing
\maketitle

\begin{abstract}
How are economic activities linked to geographic locations?
To answer this question, we use a data-driven approach that builds on the information about location, ownership and economic activities of the world's 3,000 largest firms and their almost one million subsidiaries.
From this information we generate a bipartite network of cities linked to economic activities.
Analysing the structure of this network, we find striking similarities with nested networks observed in ecology, where links represent mutualistic interactions between species.
This motivates us to apply ecological indicators to identify the unbalanced deployment of economic activities.
Such deployment can lead to an over-representation of specific economic sectors in a given city, and poses a significant thread for the city's future especially in times when the over-represented activities face economic uncertainties.
If we compare our analysis with external rankings about the quality of life in a city, we find that the nested structure of the city-firm network also reflects such information about the quality of life, which can usually be assessed only via dedicated survey-based indicators.
\end{abstract}

\section{Introduction}

The set of goods and services produced in a city depend on a complex interplay of factors that include institutions, taxes, skilled personnel, industrial heritage or the presence of particular resources. 
Dependent on the availability of such factors, some cities have specialized in certain economic activities while others became economically more diversified.
Specialization comes with a benefit as it allows for economic multiplier effects through ``agglomeration economies''~\cite{ohlin1933interregional,hoover1948location,krugman1993first}.
This, however, can turn into a drawback if a particular economic activity goes into recession.
Then, cities specialized in this activity will be distressed more than the economically diversified ones. 
To exacerbate the problem, in a globalized world the economic performance of a city increasingly depends on the economic performance of other cities.
Such economic dependencies emerge even between cities that are very far away in terms of geographical distance, which due to global economic linkages are in reality proximal in economic terms~\citep{scholl2015spatial}.

This leaves us with the problem to quantify such dependencies and to link them to the diversification of economic activities in cities. 
In this paper, we identify a city's economic activities by monitoring the firms with global presence operating in this particular city.
This way we can create a network that links cities and firms, and if we focus on the economic activities of each firm, we can extend this network by linking cities to economic activities.
Networks that describe relation between different sets of nodes, in our case cities and economic activities, are called bipartite networks (or bipartite graphs)~\cite{diestel2000graph}.
The bipartite network we construct to link every city with the economic activities of firms with global presence will allow us to study \emph{how specialized or diversified} each city is with respect to a global context. 

Analyzing the structure of this bipartite network reveals striking similarities with other types of bipartite networks found in \emph{ecology}.
There, nodes represent species and links their interactions. 
In so-called antagonistic networks, the interaction between species is asymmetric, such as host--parasite, predator--prey and plant--herbivore interactions. 
In so-called  \emph{mutualistic} networks, on the other hand, the interaction between species is symmetric, i.e. both species interact in a mutually \emph{beneficial} way like, for example, plants interact with their pollinators. 
Networks with antagonistic and mutualistic interactions have long been studied in ecology, to show that  the stability of ecological communities is linked to structural features of the network topology~\cite{bastolla:2009,Thebault2010a,rohr2014structural}. 
More precisely, it was shown that mutualistic networks are organized in a nested pattern, while antagonistic networks are organized in compartments~\cite{Thebault2010a}.
A nested organization means that the network consists of sets of generalist nodes and sets of specialist nodes. 
The specialists interact only with a small subset of nodes, while the generalists interact with (almost) all other nodes in the network. 
In nested ecosystems the large set of interactions between generalists (i.e. species that interact with many other species) creates a dense core to which the specialists (i.e. species that interact with few other species) are attached. 
It was shown that the (empirically observed~\cite{Bascompte2003}) nested structure of mutualistic networks reduces the inter-species competition, which as a consequence allows ecosystems to support more species and increase biodiversity~\cite{bastolla:2009}.

Thus, it would be of great interest to see whether such nested structures can be also found in bipartite networks related to economics.
A recent study~\cite{hidalgo2009building,tacchella2012new} has investigated the bipartite network between firms and countries, to relate it to economic stability.
It was found that robust countries have, indeed, a wide range of diversification in their economic activities. 
Based on this economic complexity performance measures for countries were proposed. 
Specifically, it was shown that the dynamics of the nested structure of industrial ecosystems can predict path dependencies in the way industries appear and disappear in given countries.
This helped to explain the evolution of the set of products that are produced and exported by these countries~\cite{Saavedra2011a,10.1371/journal.pone.0049393}.
A different analysis, focused in the New York garment industry, has shown that a firm's survival probability depends on the firm's position in the nested network of interactions between designer and contractor firms~\cite{Saavedra2011a}.

In this work, however,  we are interested in the bipartite network between cities and economic activities of globalized firms. 
These interactions are mutualistic because cities benefit from firms through taxes, employment etc, while firms benefit from cities through access to infrastructure, resources, customer base, skilled personnel etc. 
To build our bipartite network of city-economic activity relations (an extract of which is shown in Fig.~\ref{fig:GRS-1A}), we use data about firms with global presence. 
For these firms we know the precise locations of the headquarters and all their subsidiaries and, using the standard {\it Nomenclature of Economic Activities} (NACE), we classify their core business to an economic sector (for details see Data and Methods).
Next, we study the structure of this network, and we show that it follows a nested assembly, similar to ecological mutualistic networks. 
Therefore, building upon previous works in the field of ecology and their follow-ups with respect to economic networks, we apply ecological indicators to
identify the unbalanced deployment of economic activities, and we provide evidence that the structure of this bipartite network of city-firm relations contains information about the quality of life in cities.

\section{Data and Methods}

\begin{figure}[t]
  \centering
  \includegraphics[width=0.75\textwidth]{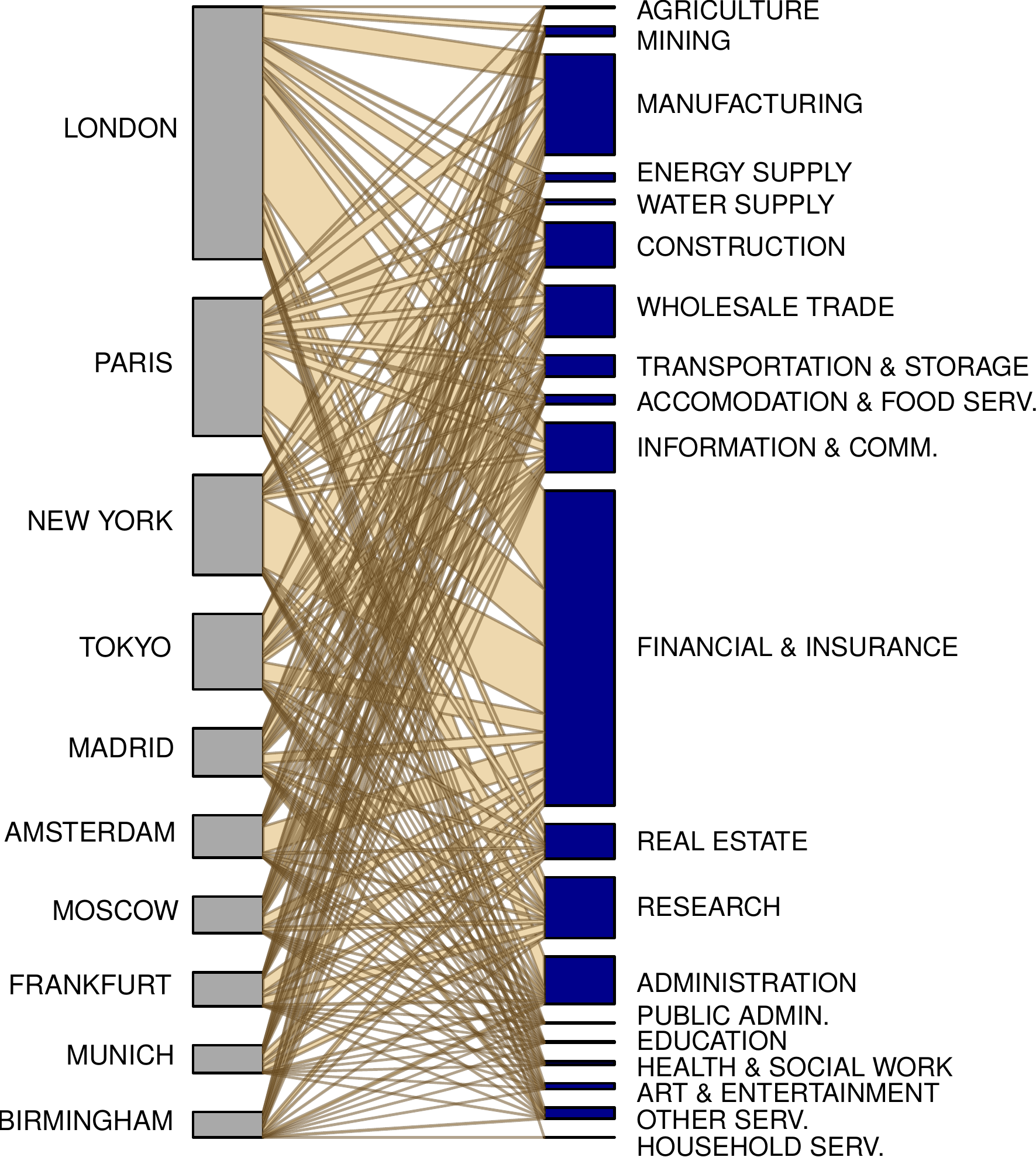}
  \caption{{\bf City - economic activity mutualistic interactions.}  City - activity interactions as a bi-partite
    network. The activity links of the 10 cities with the largest
    number of firms are shown. The link width between city $i$ and
    activity $j$ corresponds to the number of firms associated to
    activity $j$ located in city $i$.}
  \label{fig:GRS-1A}
\end{figure}

For our analysis we use data about the 3,000 largest firms with global presence
and their $\sim$1 million direct and indirect links to $\sim$800,000
subsidiaries extracted from the BvD {\it orbis} database of
2010~\cite{BVD}. The firm locations were aggregated using the concept
of Functional Urban Areas (FUA) which was developed by the {\it
  European Spatial Planning Organization Network}
(ESPON)~\cite{ESPON}. FUA's allow to agglomerate municipalities
according to their functional orientation -- sometimes going beyond
administrative boundaries -- and reflect the actual operational
conditions of people, enterprises, and communities. Therefore FUA
agglomerations result in an efficient mapping of the economic
activity and service production. In addition, we classified the firms
to economic sectors according to their core business, using the {\it
  Nomenclature of Economic Activities} (NACE) provided by
\emph{eurostat}~\cite{NACE}. These sectors were further aggregated following the United Nation's {\it International Standard Industrial Classification of All Economic Activities} (ISIC REV.4) methodology, which results to an aggregation of activities into 21 different sections.
Pairing the geographic location given by
the FUA, with the NACE-ISIC classification of every individual firm, we create a
bipartite network of interactions between 1,169 cities and 21 economic
activities. An example of this network for the 10 cities with the largest number of firms is shown in Fig.~\ref{fig:GRS-1A}.

Our bipartite network is represented by an incidence matrix $M$ pairing each of the 21 economic activities to each of the 1,169 cities where this activity is present (see Fig.~\ref{fig:GRS-1B}).
Therefore, each matrix element $m_ij$ has value one if the economic activity $i$ is present in the city $j$ and zero otherwise.  

To calculate the nestedness value of this matrix, we used the NODF algorithm developed by Almeida-Neto {\it et al}.~\cite{Almeida-Neto2008}.
This algorithm returns a nestedness value $N$ in the range [0, 100], with $N=0$ when there is no nestedness and $N=100$ for the case of perfect nestedness.
To assess the significance of nestedness we have to compare our measured value with a benchmark null model. 
In this paper, our model of choice is the null model introduced by Bascompte {\it et al}.~\cite{Bascompte2003}, which creates randomized networks by preserving the degree distribution of the original network.
This model generates ensembles of swapped matrices $\tilde{M}$ with the probability of each matrix cell to be occupied is the average of the
probabilities of occupancy of its row and column. 
Practically, this means that the probability of drawing an interaction is
proportional to the level of generalization (degree) of both the
city and the economic activity, i.e. $p_ij=(k_i/n_j+k_j/n_i)/2$, where $k_j$ is the degree of the city and $k_i$ the degree of the activity in the bipartite network, while $n_i$ and $n_j$ are the number of available activities and cities respectively. 

To measure the contribution of each individual city to the nestedness value of the whole network, we follow the methodology of Saavedra et al.~\cite{Saavedra2011a}. 
More precisely, we calculate
$c_i = (N - \mean{N_{i}^{*}})/\sigma_{N_{i}^{*}}$; were $N$ is the
observed nestedness of the whole network, $\mean{N_{i}^{*}}$ and
$\sigma_{N_{i}^{*}}$ are the average and standard deviation of the
nestedness across an ensemble of 100 random replicates for which
all the links of city $i$ to economic activities have been randomized. 
The number of random replicates is chosen in order to provide optimal performance, while at the same time the individual contribution to nestedness of each city has converged significantly to their asymptotic value. 
More precisely, we performed a convergence analysis for which we calculated the Spearman's $\rho$ correlation coefficient between two consecutive  rankings of cities according to their nestedness contribution with increasing number of replicates.
From this analysis we a) observed that the rankings indeed converge to a saturation level and b) we concluded that 100 random replicates are enough as $\rho$ is already almost 0.99.

\section{Results}

\begin{figure}[t]
  \centering
  \includegraphics[width=0.75\textwidth]{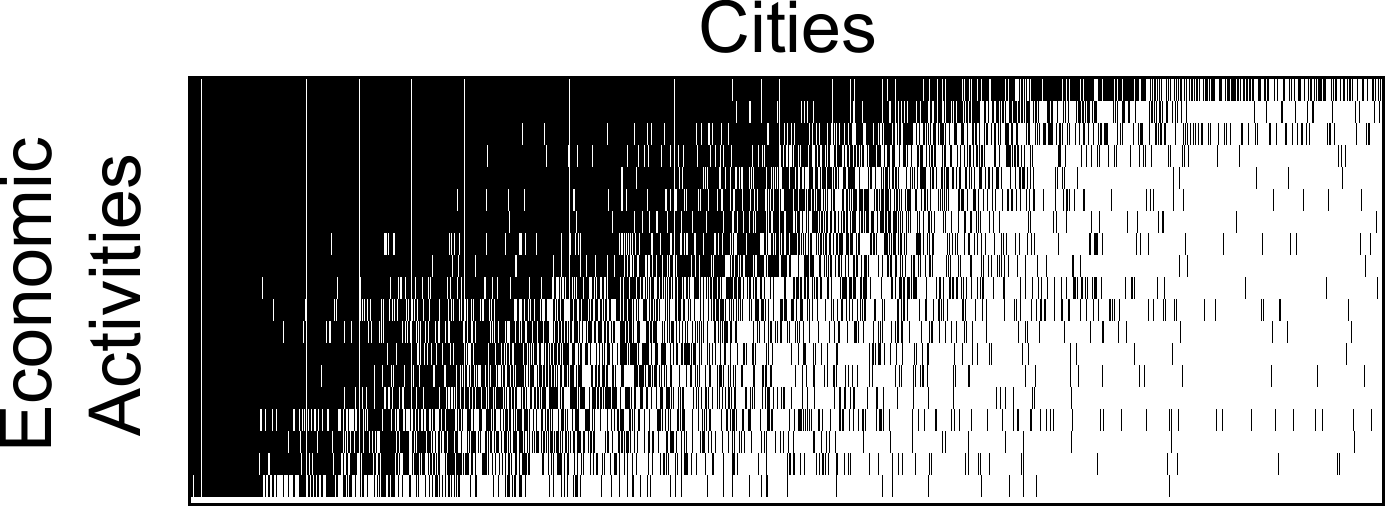}
  \caption{{\bf The city - activity interaction matrix} Plot of the interaction (incidence) matrix $M$ pairing each of the 21 economic activities to each of the 1,169 cities where this activity is present. Each matrix element $m_ij$ has value one if the economic activity $i$ is present in the city $j$ and zero otherwise.}
  \label{fig:GRS-1B}
\end{figure}

\begin{figure}[t]
  \centering
  \includegraphics[width=0.65\textwidth]{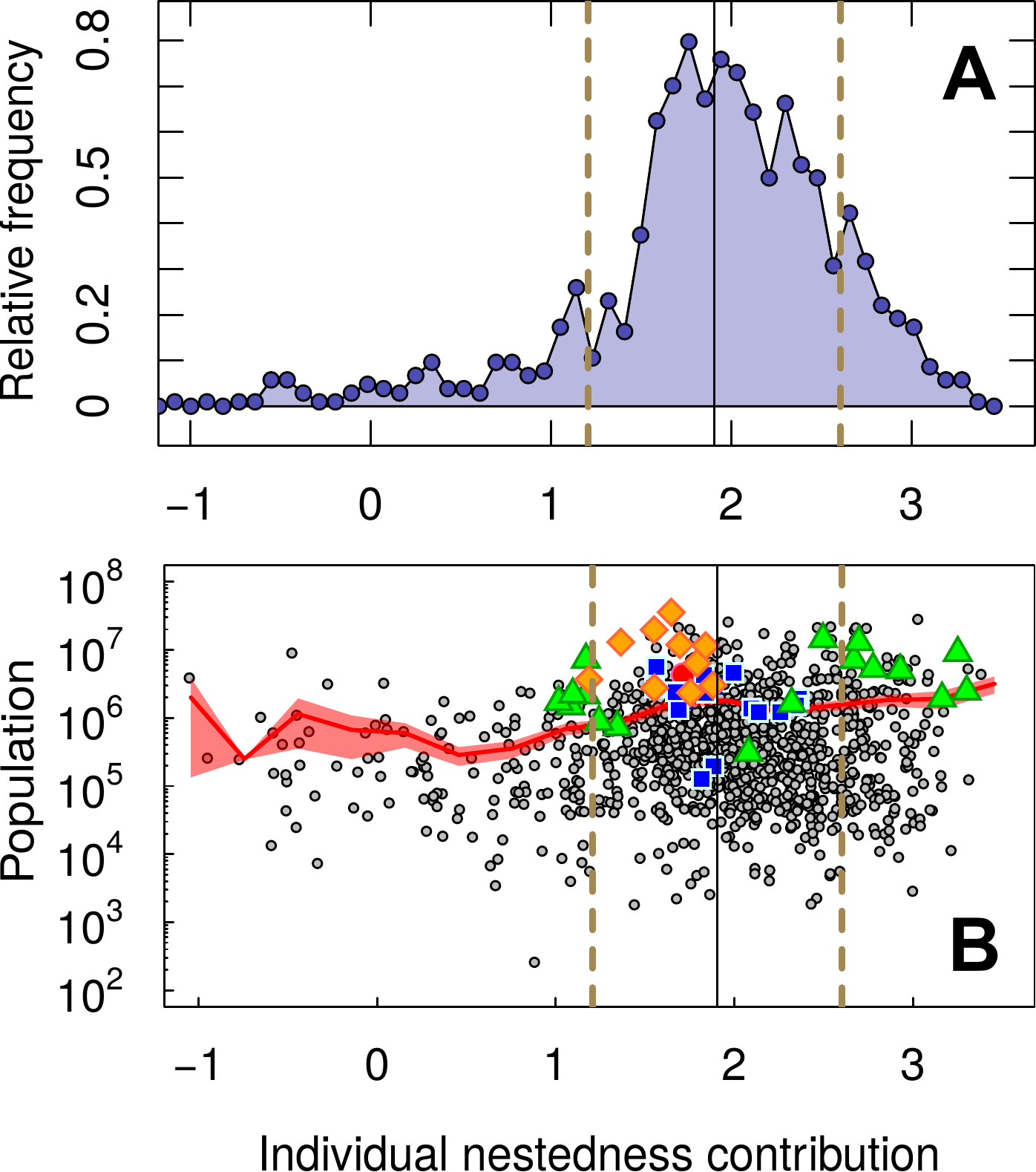}
  \caption{{\bf Individual nestedness
    contribution.}  ({\bf A}) Distribution of individual nestedness
    contribution for all cities. ({\bf B}). Population versus
    individual nestedness contribution. The red line shows the mean
    value and the band the standard error. The blue squares highlight
    the location of the top 10 cities according to the Mercer ``{\it
      2012 Quality of Living worldwide city
      rankings}''~\cite{Mercer2012} and the top 10 cities according to
    the Economist Intelligence Unit's (EIU) 2013 ``{\it Global
      Liveability Ranking and Report}''~\cite{Economist2013} while the
    green triangles highlight the location (where available) of the
    bottom 10 cities of the above rankings. In addition, with yellow
    diamonds the locations of the 10 cities with the largest number of
    firms are shown..}
  \label{fig:GRS-1C}
\end{figure}

\begin{figure}[t]
  \centering
  \includegraphics[width=0.75\textwidth]{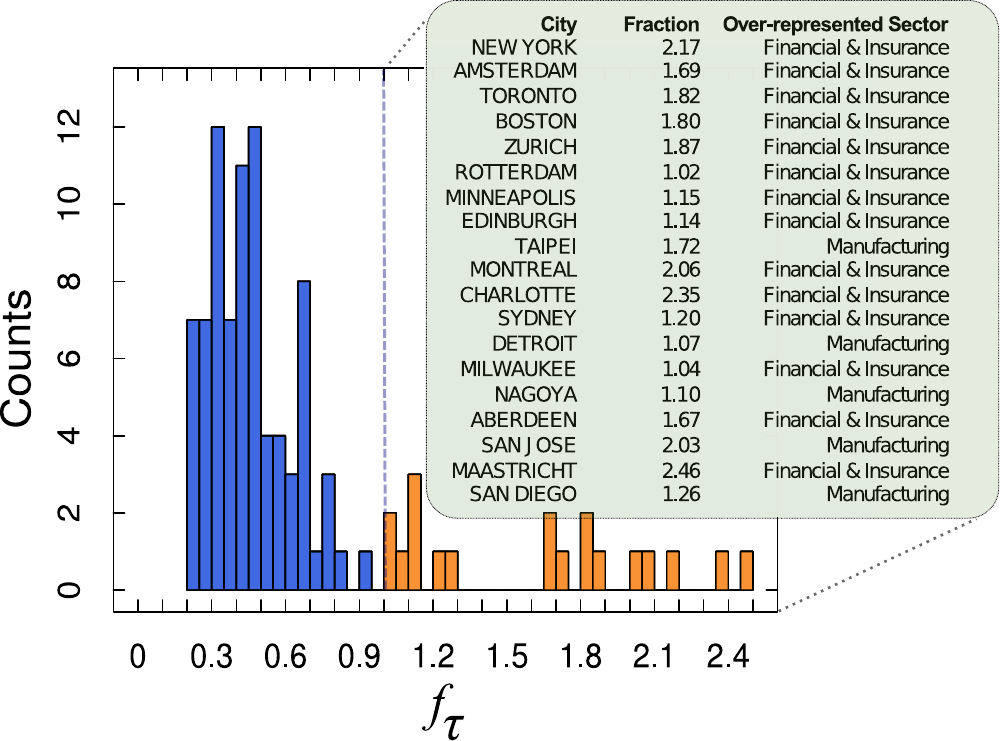}
  \caption{{\bf Concentration of economic activities.} Histogram of the fraction, $f_{\tau}$, for the 100 cities
    with the largest number of firms in our database.}
  \label{fig:GRS-2}
\end{figure}

Using the 
methodology described above, we find that the bipartite network of cities-economic activities
is nested (see Fig.~\ref{fig:GRS-1B}), with a
nestedness value $N=0.784$ ($\mathrm{p}<0.0001$). 
This already highlights structural similarities in the interaction patterns that occur in a natural ecological system and in the human made economic system. 
And since a nested network structure is known to promote community stability in mutualistic ecological networks~\cite{Thebault2010a}, we anticipate that the mutualistic network of cities and economic activities would be stable as well.

But, as it was shown recently both for ecological and socio-economic
networks, nestedness comes with a price~\cite{Saavedra2011a}. The nodes
that contribute more to the nestedness of the network are the nodes that
contribute more to the network persistence. However, these same nodes were
identified as the most vulnerable to go extinct. Of course in our case
a city may not go extinct, but it may decline to a less prosperous
state.

As shown in
Fig.~\ref{fig:GRS-1C}A, the distribution of the individual contribution to nestedness is concentrated around the
mean value $\mu=1.96\pm 0.01$.  
Therefore, it is an important question whether this nestedness value has any relation to a city's economic performance.
If it does so, then where are the best performing cities located in this distribution? close to the center or close to its tail?

Unfortunately, we do not have access to data about economic performance for individual cities. 
However, we do expect economic performance to be strongly correlated with the well being of a city's inhabitants. 
In this sense, using the rankings provided by the Economist Intelligence Unit's
(EIU) 2013 ``{\it Global Liveability Ranking and
  Report}''~\cite{Economist2013} and the Mercer ``{\it 2012 Quality of
  Living worldwide city rankings}''~\cite{Mercer2012}, we calculated
the nestedness contribution of the top-10/bottom-10 performing cities.
We found that all of the top cities of the above rankings are within
the range of $\mu \pm \sigma$, while 70\% of the bottom cities are
outside this range. More precisely, 41\% of them are above $\mu +
\sigma$, and 39\% below $\mu - \sigma$. From Mercer's bottom-10 list
the cities above $\mu + \sigma$ are Abidjan, Khartoum, Kinshasa,
Conakry, while from EIU's list Karachi, Algiers, Douala are in this
range. In the area below $\mu - \sigma$ from Mercer's bottom-10 list
we find Tbilissi, Sanaa, Baghdad, and from EIU's list Damascus and
Tripoli.
This discussion shows that a ranking based on the nestedness score
gives insightful results, where the better performing cities,
according to Mercer and Economist~\cite{Mercer2012,Economist2013} are
closer to the mean of the nestedness distribution, while the worst
performing ones are further away (Fig.~\ref{fig:GRS-1C}B).

Given the general tendency of cities to grow, it is natural to ask if there is any measurable impact of
population to the nestedness score.  
It is known already that a city's population drives many diverse properties of cities~\cite{bettencourt2007growth}.
Are smaller cities more stable or more vulnerable according to the way stability/vulnerability  is reflected through
nestedness? To answer this question we collected data about city population,
by consolidating information based on the UN database on cities\footnote{\url{http://data.un.org}}, the OECD database on cities\footnote{\url{https://stats.oecd.org/Index.aspx?DataSetCode=CITIES}}, and the ESPON
project~\cite{ESPON}. 
As shown in Fig.~\ref{fig:GRS-1C}B, there is no
pronounced relation between the (logarithm of) population and the
individual nestedness. The Pearson correlation coefficient $r=0.069$ ($\rm
p=0.017$) is small and not significant, and the same holds true for the Spearman correlation $\rho=-0.0052$ ($\rm
p=0.8588$).  Of course, if a city performs
well and increases its inhabitants' well being, it may become the
target of internal or external migration flows and eventually increase
its population. However, its network position --as measured with
respect to nestedness-- does not seem to be influenced by the population.

We can of course anticipate that if a city is specialized in an economic 
activity, it will prosper as long as the activity fares well. If this
activity is hit by turmoil, or just under-performs with respect to other activities, this may lead to a
decline of the city. To avoid such risks, diversification of
activities is required; but how much diversification is enough? And
even if a city has indeed diversified its activities, how does this
diversification compare to other cities?
It is expected that large cities are able to attract many firms, that
would populate multiple economic sectors of activity~\cite{pumain2006evolutionary}. This means that
large cities are by definition ``generalists'' in the bipartite graph,
and this introduces a bias in our interpretation of the nestedness
score. 

To be more specific, let us consider the case of Detroit with
nestedness score $c=1.71$, which places it near the mean of the
nestedness distribution. Based on this number alone we would argue
that Detroit performs well and we would not anticipate its bankruptcy
on July 18, 2013. Therefore, it is not
enough to only ensure that multiple economic sectors are populated,
but, it is important to monitor how many firms populate each sector.
If the distribution of firms in economic sectors is skewed, one or few
sectors will be dominating. So, a major decline in the dominating
sector will have a major impact on the city's economy, and this will
indirectly affect all the other sectors, as well.

In addition, this effect will be even more pronounced in the presence of hidden links between firms from different sectors.
I.e. it is not hard to imagine that many service related smaller firms (e.g. subcontractors or advance production services) shall provide support and will be dependent to the function of the large firms of the dominant economic sector.
Therefore, the decline of this sector will create a cascading effect that is very hard to be properly evaluated in the absence of detailed dependency data. 

In the example at hand, from the 4,455 total large firms that were
active in Detroit, 2,299 belonged to the manufacturing sector. The
second most populated sector was Financial \& Insurance with 642
firms. We expect that many of these firms have strong ties to the manufacturing companies, and shall be affected if something will go wrong in the manufacturing sector. 
However, since we cannot document these ties, for simplicity we will assume that all sectors are independent. 

Hence, to detect when one sector is over-represented in the
overall economic activity we calculate the fraction $f_{\tau}$,
i.e. the number of firms in the largest sector over the number of
firms in all other sectors. 
If $f_{\tau}\leq 1$ it would mean that the city is well diversified across activities, while if $f_{\tau}>1$ a particular sector is dominating the economy, and the city might be at risk.
For our dataset, and under the  assumption of sectoral independence, $f_{\tau}= 1.066$  for the case of Detroit, which indicates the city's vulnerability.   

Provided the availability of more refined data, we could improve the calculation by dividing the number of firms of all sectors that are significantly affected by a decline in the largest sector by the number of firms that will not be affected by this decline.
However, such calculation cannot be performed with our currently available datasets.

We would like to contrast our measure  $f_{\tau}$ with other existing indexes for diversity. 
A well known index in Ecology  to quantify the biodiversity of a habitat is the Simpson's index~\cite{simpson1949measurement}. 
It is also known as Herfindahl-Hirschman index in economics, where it is used to measure market concentration~\cite{hirschman1980national,herfindahl1950concentration}.
We calculated the Simpson's index for all cities in our database, and we found that the resulting concentration ranking is strongly correlated (Spearman's $\rho=-0.976$ ($\rm p=<0.0001$)) with the ranking based on $f_{\tau}$.
But most available indexes including the Simpson's index do not allow to easily identify a threshold value that discriminates well diversified cities from not well diversified cities. 
This, however, can be achieved by the $f_{\tau}=1$ value in our case. 

There is a limitation when applying our $f_{\tau}$ index to extremely specialized cities, as it diverges in cases where (mostly due to data limitations) only one economic sector is present. 
These cases are identified by the Simpson's index as extremely specialized cities, as well, and are assigned a zero value.
It is, therefore, better practice to exclude such pathological cases from our analysis. 
For this reason we restricted our calculation of $f_{\tau}$ to the 100 cities with the largest number of firms in our database~\footnote{We calculated the Simpson's index for the set of 100 cities with the largest number of firms in our database, and again we found that the resulting concentration ranking is strongly correlated (Spearman's $\rho=-0.963$ ($\rm p=<0.0001$)) with the ranking based on $f_{\tau}$.}. 

As shown in
Fig.~\ref{fig:GRS-2}, most of the cities have a $f_{\tau}$ smaller
than one, which is evidence of a balanced development. However, there
are some cities with $f_{\tau}$ values not only larger than one, but
even larger than the value of Detroit. As it happens most of these cities, which include New York, Amsterdam, Zurich etc. are large
financial centers, which highlights the fragility of an economic model
that is largely dependent on financial services. The recent financial
crises rang some bells, and now policy makers in developed countries
try to mitigate this dependency by re-shoring the manufacturing
sector. A profound example is EU $10|100|20$ strategy, which aims to
get almost 20\% of semiconductor manufacturing back to Europe by 2020
through an unprecedented public/private investment
partnership\footnote{\url{http://www.semi.org/eu/node/8506}}.

\section{Conclusion}

In summary, by exploiting functional similarities across complex
systems, we can use indicators developed in ecology to assess the
performance of a city in the globalized economy.
With these indicators we go beyond the mere evaluation of the economic
specialization of cities~\cite{isard1956}, as we associate
specialization of one city to the vulnerability of the whole ``ecosystem'' describing city - economic activities relations, similar to the way extinction of one species affects the stability of natural ecosystems.
It is shown that such indicators have the potential to identify the need for
new multilevel policies, able to regulate the cities at the 
national or continental level (like within EU), in order to enhance their position in the bipartite network of city - economic activities relations. 

However, there is also a need for closer supervision to prevent
over-representation of some economic activities at the expense of
others, as this increases the risk in the future. In this respect,
policy interventions that reduce the dominance of one sector over the
others should be applied more frequently. Currently, the financial
sector is strongly over-represented in most of the large cities of the
developed countries, hence, policies like the EU $10|100|20$ strategy
are important as hedging against future risks.

\section*{Acknowledgment}

The authors acknowledge financial support from the EU-FET project
MULTIPLEX 317532. We thank Antoine Bellwald and Faraz Ahmed Zaidi for
cleaning and aggregating the data. For our analysis we used the R
software for statistical analysis v3.0.2, and the bipartite library
v2.02.

\bibliography{GRS}
\bibliographystyle{sg-bibstyle}

\end{document}